\documentclass[aps,twocolumn,prd,nofootinbib,superscriptaddress,preprintnumbers]{revtex4-2}

\usepackage[utf8]{inputenc}

\usepackage{mathtools}
\usepackage{amsfonts}
\usepackage{amssymb}
\usepackage{mathrsfs}
\usepackage{bbm}
\usepackage{braket}
\usepackage{slashed}
\usepackage{tensor}

\usepackage{graphicx}
\usepackage{array}

\usepackage{placeins}
\usepackage{makecell}
\usepackage[caption=false]{subfig}
\usepackage{float}

\usepackage{xspace}
\usepackage{xfrac}
\usepackage{hyperref}
\usepackage[nameinlink]{cleveref}
\usepackage{appendix}
\usepackage{units}

\usepackage{xifthen}
\usepackage{booktabs}
\usepackage{multirow}
\usepackage{courier}
\usepackage[dvipsnames]{xcolor}
\hypersetup{
	colorlinks,
	linkcolor={blue!75!black},
	citecolor={blue!75!black},
	urlcolor={blue!75!black}
}

\Crefname{equation}{Eq.}{Eqs.}

\let\Re\undefined
\DeclareMathOperator{\Re}{Re}
\DeclareMathOperator{\Tr}{Tr}

\graphicspath{{./figs/}}

\newcommand{\gettitle}{Applications of flow models to the generation of correlated lattice QCD ensembles}

\newcommand{\getMITAffiliation}{\affiliation{Center for Theoretical Physics, Massachusetts Institute of Technology, Cambridge, MA 02139, USA}}
\newcommand{\getIAIFIAffiliation}{\affiliation{The NSF AI Institute for Artificial Intelligence and Fundamental Interactions}}
\newcommand{\getFNALAffiliation}{\affiliation{Fermi National Accelerator Laboratory, Batavia, IL 60510, U.S.A.}}
\newcommand{\getDMAffiliation}{\affiliation{Google DeepMind, London, UK}}
\newcommand{\getBernAffiliation}{\affiliation{Albert Einstein Center, Institute for Theoretical Physics, University of Bern, 3012 Bern, Switzerland}}

\hypersetup{
    pdftitle={\gettitle},
    pdfauthor={Abbott,Boyda,Hackett,Kanwar,Racanière,Rezende,Romero-López,Shanahan,Urban},
    pdfkeywords={lattice field theory}{lattice gauge theory}{normalizing flows}{machine learning},
    bookmarksopen=true,
    bookmarksopenlevel=2,
    bookmarksnumbered=true
}

\begin{document}

\title{\gettitle}

\author{Ryan~Abbott}
\getMITAffiliation
\getIAIFIAffiliation
\author{Aleksandar Botev}
\getDMAffiliation
\author{Denis~Boyda}
\getMITAffiliation
\getIAIFIAffiliation
\author{Daniel~C.~Hackett}
\getFNALAffiliation
\getMITAffiliation
\getIAIFIAffiliation
\author{Gurtej~Kanwar}
\getBernAffiliation
\author{S\'{e}bastien~Racani\`{e}re}
\getDMAffiliation
\author{Danilo~J.~Rezende}
\getDMAffiliation
\author{Fernando~Romero-L\'opez}
\getMITAffiliation
\getIAIFIAffiliation
\author{Phiala~E.~Shanahan}
\getMITAffiliation
\getIAIFIAffiliation
\author{Julian~M.~Urban}
\getMITAffiliation
\getIAIFIAffiliation

\preprint{MIT-CTP/5658,\,FERMILAB-PUB-24-0014-T}

\begin{abstract}
    Machine-learned normalizing flows can be used in the context of lattice quantum field theory to generate statistically correlated ensembles of lattice gauge fields at different action parameters.
    This work demonstrates how these correlations can be exploited for variance reduction in the computation of observables.
    Three different proof-of-concept applications are demonstrated using a novel residual flow architecture: continuum limits of gauge theories, the mass dependence of QCD observables, and hadronic matrix elements based on the Feynman-Hellmann approach.
    In all three cases, it is shown that statistical uncertainties are significantly reduced when machine-learned flows are incorporated as compared with the same calculations performed with uncorrelated ensembles or direct reweighting. 
\end{abstract}

\maketitle

\section{Introduction}\label{sec:intro}

Understanding the strongly interacting sector of the Standard Model of particle physics, described by the theory of quantum chromodynamics (QCD), is essential for advancing particle and nuclear physics.
The numerical framework of lattice QCD is a systematically improvable tool to explore the dynamics of the strong nuclear force.
This approach has enabled precise calculations across applications spanning from hadron structure to high-temperature QCD and nuclear physics~\cite{Boyle:2022ncb, USQCD:2022mmc}.
Nevertheless, there is great potential to extend the reach of lattice QCD beyond the current state of the art if computational challenges such as critical slowing down, topological freezing, and signal-to-noise problems can be overcome.
In this context, emerging machine learning techniques offer a promising avenue towards mitigating these computational obstacles~\cite{Boyda:2022nmh, Cranmer:2023xbe}. 

A growing community effort is developing at the intersection of machine learning and lattice QCD---see e.g. Refs.~\cite{Bachtis:2021xoh, Cali:2022qbd, Lehner:2023prf, Wang:2023exq, Nicoli:2020njz} for a selection of applications.
In particular, generative flow models~\cite{rezende2016variational, dinh2017density, JMLR:v22:19-1028} are one of several promising pathways which show potential to accelerate the sampling  of lattice field configurations.
This line of investigation is developing, with demonstrations in 2D theories~\cite{LiWang2018NNRG, Albergo:2019eim, Kanwar:2020xzo, Boyda:2020hsi, Hackett:2021idh, Albergo:2021bna, Albergo:2021vyo, Albergo:2022qfi, Nicoli:2020evf, Nicoli:2020njz, Foreman:2021ixr, Foreman:2021ljl, Foreman:2021rhs, DelDebbio:2021qwf, Gabrie:2021tlu, deHaan:2021erb, Lawrence:2021izu, Jin:2022bgq, Pawlowski:2022rdn, Finkenrath:2022ogg, Gerdes:2022eve, Singha:2022lpi, Matthews:2022sds, Caselle:2022acb, Albandea:2023wgd, Albandea:2023ais, Bacchio:2022vje, Nicoli:2023qsl, Singha:2023xxq} and first applications to 4D gauge theories with and without fermions~\cite{Abbott:2022zhs, Abbott:2022hkm, Abbott:2023thq}.
While the field is progressing rapidly, achieving high-quality models that can be applied at the scale of state-of-the-art calculations still requires further 
engineering~\cite{Abbott:2022zsh}.
In addition to their promise in the context of sampling, flow models---functioning as approximate maps between distributions---can be used to accelerate lattice QCD calculations in qualitatively different ways.
For example, flow models provide a promising new approach to determining thermodynamic observables~\cite{Nicoli:2020njz, Nicoli:2021inv, Pawlowski:2022rdn, Nicoli:2023qsl}.

In this work, we explore applications which utilize flows to map gauge field configurations between distributions defined by different Euclidean lattice action parameters.
Such flows can be used to generate multiple statistically correlated ensembles at different parameters.
As we explore in this work, this may be particularly valuable  when the variation of some quantity with respect to the action parameter is of physical or computational interest---see also Refs.~\cite{Bacchio:2023all,Catumba:2023ulz}.
The advantage of flows in this context originates from correlated cancellations of uncertainties between expectation values evaluated at different action parameters, which leads to reductions in the number of configurations needed to achieve a fixed statistical error.

Examples of physically relevant applications of derivatives with respect to action parameters include continuum and chiral extrapolations as well as the computation of matrix elements such as the chiral condensate, the nucleon sigma term, or other observables, using Feynman-Hellmann techniques.
Another is derivatives with respect to the electromagnetic coupling for scale setting or to compute isospin breaking corrections in QCD+QED~\cite{deDivitiis:2013xla, Tantalo:2023onv}.
One may also consider applications in theories with a sign problem, e.g., to 
derivatives with respect to the baryon chemical potential or the QCD $\theta$-term.
In all of these cases, the distributions to be related by a flow transformation are much more similar than in applications intended to accelerate sampling, and current flow methods can already be applied at the scale of typical lattice QCD calculations.
Three selected applications are investigated, namely the continuum extrapolation of gradient flow scales, the computation of the gluon momentum fraction of the pion in quenched lattice QCD using the Feynman-Hellmann approach, and the mass dependence of observables in $N_f=2$ QCD. 

This paper is organized as follows.
In \Cref{sec:corrflows}, we discuss preliminaries on flows, their applicability in the context of correlated ensembles, and the residual flow architectures used in this work.
The three numerical demonstrations are presented in \Cref{sec:demos}. We conclude in \Cref{sec:conclu}. \Cref{app:arch} provides further details of the flow models used in this work.

\section{Flows for the generation of correlated ensembles}\label{sec:corrflows}

\subsection{Flows for lattice QCD}\label{subsec:flows}

This section presents an introduction to normalizing flows~\cite{rezende2016variational, dinh2017density, JMLR:v22:19-1028}, reviewing the key ideas relevant for the present work.

A ``flow'' is defined as a diffeomorphism $f$ between probability distributions that maps samples from a base (or prior) distribution, $r(U)$, to a model distribution with density 
\begin{equation}
    q(V) = r(U)  \left|\det \frac{\partial f(U)}{\partial U}\right|^{-1} \ ,
\end{equation}
where $V=f(U)$. Flows can be constructed such that they have many free, trainable parameters.
These parameters may be optimized such that the model distribution approximates some target distribution $p$, i.e., ${q(V)\simeq p(V)}$.

For the applications explored in this work, flow models are constructed in which the samples $U$ are lattice gauge-field configurations, and the probability distributions $p(U)$ and $r(U)$ are defined in terms of Euclidean lattice actions such that $r(U) \propto \exp (-S_0(U))$, and $p(V) \propto \exp (-S_1(V))$.
In most cases, it is not necessary to know the normalization of $p$ or $r$ (the exception being thermodynamic observables~\cite{Nicoli:2020njz}). 

Expressive flow transformations can be constructed in a variety of ways, for example as the composition of 
$n$ invertible layers
\begin{equation}
    f = g_1 \circ g_2 \circ ... \circ g_n \ .
\end{equation}
Architectures for invertible layers $g_i$ which act on lattice gauge fields have been discussed in Ref.~\cite{Abbott:2023thq}.
The particular constructions used in this work are detailed in \Cref{subsec:arch}.
Given a model, its trainable parameters may be optimized in various ways.
One choice is to minimize the Kullback-Leibler (KL) divergence~\cite{Kullback:1951} between the model and target distributions.
Approaches such as path gradients~\cite{vaitl2022gradients}, related control variate methods~\cite{Abbott:2023thq}, as well as the ``REINFORCE'' algorithm~\cite{Bialas:2023fyj}, may be be used to improve and accelerate training dynamics by reducing the variance associated with stochastic gradient estimates.
After optimization, model quality can be characterized using the Effective Sample Size per configuration (ESS),
\begin{equation}\label{eq:ESSgeneral}
    \text{ESS } = \frac{1}{N} \frac{\left[ \sum_{i=1}^N w(V_i)  \right]^2}{\sum_{i=1}^N \big [w(V_i)\big]^2} \ ,
\end{equation}
estimated using $N$ gauge field configurations generated from $q(V)$, and where $w(V_i) = p(V_i)/q(V_i)$ is the reweighting factor of the $i$th configuration.
The values of the ESS lie in the interval ${\text{ESS} \in [1/N\, ,\, 1]}$, with ${\text{ESS}=1}$ corresponding to a perfect model.

In practice, a learned flow is not perfect, but may function as an approximate map between distributions.
To ensure correctness of expectation values computed on the flowed configurations, one may use the independence Metropolis algorithm~\cite{Metropolis:1953am, Hastings:1970aa, tierney1994markov} or simply reweighting, with the weight of each configuration given by $w(U)$.
Expectation values of observables such as plaquettes, hadronic correlation functions, or the topological charge can be directly reweighted as
\begin{equation}
    \langle \mathcal O \rangle_p = {  \langle w \, \mathcal O \rangle_q } \ ,
\end{equation}
where the notation $\langle \rangle_q$ is used to refer to expectation values with respect to the probability distribution $q$, and we assume the reweighting factors have been properly normalized such that ${ \langle w \rangle_q}=1$.
Derived quantities, such as gradient flow scales or hadron masses, can be computed from reweighted correlation functions. Statistical uncertainties in reweighted quantities are typically larger than those before reweighting. A rough estimate of the increase in the variance is a factor of $\simeq 1/\text{ESS}$. 

\subsection{Correlated ensembles and flows}\label{subsec:gencorr}

While applications of flows to accelerate the generation of field configurations continue to advance, here we describe another avenue for flow models to improve lattice QCD calculations by reducing the variance of observables that can be computed from differences between quantities at different action parameters. The key idea is the following. Consider a generic parameter of the action, $\alpha$.
The goal is to compute some observable $\mathcal O$ as a function of $\alpha$, and in particular the derivative
\begin{equation}\label{eq:derOalpha}
    \frac{d \langle \mathcal O \rangle}{d \alpha} \simeq \frac{\langle \mathcal O \rangle_{\alpha_1} - \langle \mathcal O \rangle_{\alpha_2}}{\Delta \alpha} \ ,
\end{equation}
where the right-hand side is a finite-difference approximation of the derivative using $\Delta \alpha=\alpha_1 - \alpha_2$, with $\langle \rangle_\alpha$ denoting the expectation under the distribution defined by the action parameter $\alpha$, i.e., $p_\alpha$.
Higher order derivatives, or derivatives of one observable with respect to another, may be computed in a similar way.

In this work, we consider three qualitatively different approaches to the computation of the quantity in \Cref{eq:derOalpha}.
The first two are standard tools in common use:
\begin{enumerate}
    \item Use a very small step $\Delta \alpha = \epsilon$, and compute the numerator in \Cref{eq:derOalpha} with {\bf $\mathbf \epsilon$ reweighting}:
    \begin{equation}\label{eq:eps-reweighting-def}
        \langle \mathcal O \rangle_{\alpha_1} - \langle \mathcal O \rangle_{\alpha_1+\epsilon} = \langle  \mathcal O - w_\epsilon\mathcal O  \rangle_{\alpha_1} \ ,
    \end{equation}
    where $w_\epsilon = p_{\alpha_1 + \epsilon}/  p_{\alpha_1}$. For this approach, the ESS generically degrades as ${\text{ESS} = 1 - k (\Delta \alpha)^2} + \hdots$, where $k$ is a problem-specific constant. The separation $\epsilon$ may be made small without compromising signal-to-noise due to correlated noise cancellations between the two expectation values.
    As $\epsilon \rightarrow 0$ it becomes exact, recovering an estimate statistically identical to that obtained by applying the derivative analytically.
    \item Generate {\bf independent ensembles} to separately compute expectation values at $\alpha_1$ and $\alpha_2$ in \Cref{eq:derOalpha}.
    This enables use of much more widely separated $\alpha_1$ and $\alpha_2$ than accessible with reweighting, thereby allowing exploitation of the bias-variance tradeoff to reduce statistical uncertainties while accepting additional discretization artifacts from the finite-difference approximation in order to improve signal-to-noise.
    However, this effect must be sufficiently large to compensate for the lack of correlated noise cancellations.
\end{enumerate}
These two methods each have different capabilities, with each useful for different applications.
Incorporating flows provides an additional approach that combines some of the advantages of both:
\begin{enumerate}
    \setcounter{enumi}{2} 
    \item Use a trained {\bf flow model} to map configurations between the distributions given by $\alpha_1$ and $\alpha_2$.
    Including flow reweighting factors, correlated differences can be calculated as:
    \begin{equation}\label{eq:flow-reweighting-def}
        \left\langle \mathcal O(U) - w(f(U)) \, \mathcal O(f(U)) \right\rangle_{\alpha_1} \ ,
    \end{equation}
    where $w(f(U)) = p_{\alpha_2}(f(U))/q(f(U))$, such that a perfect flow would remove the reweighting factors entirely.
    This approach benefits from the same correlated cancellation of uncertainties as does $\epsilon$ reweighting, while allowing for larger steps in $\Delta \alpha$ to exploit the bias-variance tradeoff as does the approach using independent ensembles.
\end{enumerate}
In \Cref{sec:demos} below, we provide numerical demonstrations of the advantages of this flow-based approach.

Note that the latter two approaches, with finite separation in $\alpha$, can be combined with improved finite-difference estimators of derivatives to reduce the $O(\Delta \alpha)$ bias, or by fitting the $\alpha$ dependence at the cost of introducing model dependence.

\subsection{Architecture based on residual flows}\label{subsec:arch}

The flow architecture used in this work is based on that introduced in Ref.~\cite{Abbott:2023thq}, with a series of improvements that are detailed below.
The flow transformation is defined as the composition of trainable gauge-equivariant layers that act directly on the gauge links.
The transformation of a gauge field $U\rightarrow U'$ through an $\mathrm{SU}(N)$-residual layer can be expressed as
\begin{equation}
    U'_\mu(x) = e^{g_x(U)} U_\mu(x) \ ,
\end{equation}
where $g_x(U)$ is an algebra-valued matrix which can in principle have an arbitrary dependence on the entire gauge-field configuration, as long as it transforms locally under gauge transformations, $g_x(U) \to \Omega^\dagger_x g_x(U) \Omega_x$; here $\Omega_x$ denotes a gauge transformation and the subscript labels the spacetime dependence.
This transformation can be inverted by fixed point iteration, 
with a unique solution guaranteed if the Lipschitz continuity condition is satisfied~\cite{Abbott:2023thq}. 

\begin{figure}
    \includegraphics[width=\linewidth]{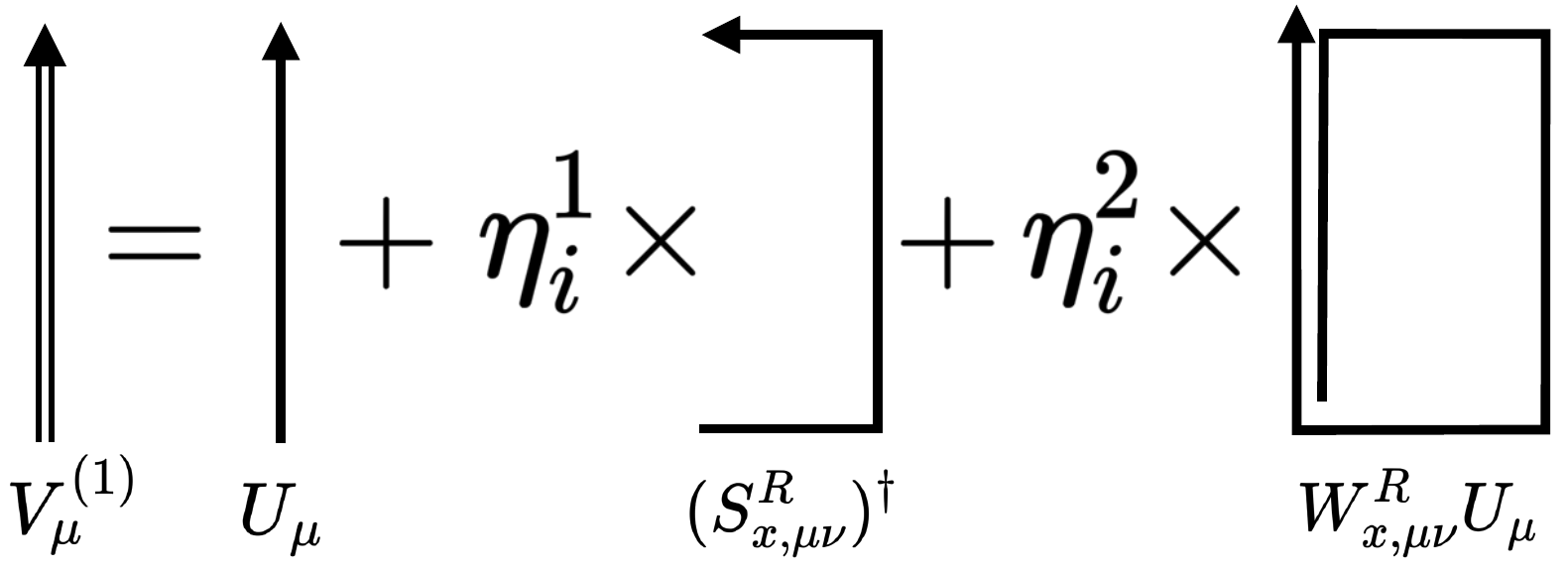}
    \caption{ Sketch of the recursive transformation, \Cref{eq:PTconv}, to build generic Wilson loops in the residual layers.}
    \label{fig:archsketch}
\end{figure}

For numerical tractability, each layer partitions the gauge field and transforms only the \textit{active links}, defined as those with fixed direction $\mu$ on a subset of lattice sites $\{ x_a\}$, conditioned on the values of the remaining \emph{frozen links} $U_f$.
Each layer acts as
\begin{equation}\label{eq:reslayer}
    U'_\mu(x_a) = e^{g_x(U_f, U_\mu(x_a))} U_\mu(x_a) \ ,
\end{equation}
that is, $g_x$ for any given active link depends on all frozen links but only the same active link.
This separation of variables allows efficient computation of the Jacobian of the transformation using automatic differentiation as described in Eq.~(26) of Ref.~\cite{Abbott:2023thq}.
In the present work, we use two partitioning schemes (also referred to as ``masking patterns'') for the site index:
\begin{enumerate}
    \item A checkerboard or  ``mod 2'' masking pattern, where the active links are those with direction $\mu$ in the positions that satisfy ${(p+\sum_\mu x_\mu) = 0\; (\mathrm{mod} \, 2)}$ for for $p \in {0,1}$. A stack of 8 layers is needed to transform all links, i.e., 2 complementary checkerboards in each of the 4 directions $\mu$. This is a simple nontrivial choice that updates all variables within a small number of layers. 
    \item A ``mod 4'' masking pattern, where the positions of active links satisfy ${(p+\sum_\mu x_\mu) = 0\; (\mathrm{mod} \, 4)}$, for $p \in {0,1,2,3}$. 16 layers are thus needed to transform every link on the lattice. This choice is more expensive than the ``mod 2'' pattern described above, but it can also be more expressive by allowing a more complicated dependence of the transformation on the frozen links.  
\end{enumerate}
The function $g_x(U_f, U_\mu(x_a))$ must be constructed in a way that is expressive but simple to evaluate.
One simple construction utilizes $1\times 1$ staples (depicted in \Cref{fig:archsketch}),
\begin{equation}\label{eq:staples}
\begin{aligned}
     S^R_{x,\mu\nu}(U) &= U_\nu(x+\mu) U^\dagger_\mu(x+\nu)  U^\dagger_\nu(x) ~~\text{and} \\
     S^L_{x,\mu\nu}(U) &= U^\dagger_\nu(x+\mu -\nu ) U^\dagger_\mu(x-\nu)  U_\nu(x-\nu) \ ,
\end{aligned}
\end{equation}
such that the $1\times 1$ loops,
\begin{equation}\label{eq:WRandWL}
\begin{aligned}
    W^R_{x,\mu\nu}(U) &= U_\mu(x) S^R_{x,\mu\nu}(U_f) ~~\text{and} \\
    W^L_{x,\mu\nu}(U) &= U_\mu(x) S^L_{x,\mu\nu}(U_f) \ ,
\end{aligned}
\end{equation}
have the same gauge transformation as $g_x$.
One can then define a covariant algebra-valued object as, e.g.,
\begin{equation}\label{eq:Gmu}
\begin{aligned}
    {G_{x,\mu}} &= \, \sum_{\nu \neq \mu} \ {\alpha^{(1)}_{\mu \nu}}  
    \mathcal{P}( W_{x,\mu\nu}(U) ) \\
     &+ \sum_{\nu,\rho \neq \mu } {\alpha^{(2)}_{\mu \nu \rho}} \mathcal{P}( W_{x,\mu\nu}(U) W_{x,\mu\rho}(U) )
    \ ,  
\end{aligned}
\end{equation}
where $W_{x,\mu\nu} = W^R_{x,\mu\nu} + W^L_{x,\mu\nu}$, and $\mathcal{P}(W)$ is the gauge-covariant traceless anti-Hermitian projection of $W$. Moreover, $\alpha^{(1)}_{\mu \nu}$ and $\alpha^{(2)}_{\mu \nu \rho}$ are $d-1$ and $(d-1)^2$ trainable parameters in $d$ spacetime dimensions for fixed $\mu$, respectively. Any polynomial function of $G_{x,\mu}$ with coefficients that are arbitrary function of $\text{Tr}[ G_{x,\mu} G^\dagger_{x,\mu}]$ 
is thus gauge covariant and can be used to construct $g_x(U)$.
One choice of such a construction is:
\begin{equation}\label{eq:gxdef}
   g_x(U_f, U_\mu(x_a)) = G_{x,\mu} \times f \left( \Tr [G_{x,\mu} G^\dagger_{x,\mu}]   \right) \ ,
\end{equation}
where $f(x)$ is e.g.,~a ratio of polynomials---see \Cref{app:arch} for an example.

A useful modification to this construction is to consider Wilson loops that are larger than  $1 \times 1$. 
Sums of such loops can be constructed iteratively, by repeatedly adding together links and staples which transform in the same way, and finally computing a $1 \times 1$ loop.
This is inspired by similar transformations used in Refs.~\cite{Favoni:2020reg, Abbott:2022zhs} and resembles the learned smearing of Ref.~\cite{Tomiya:2021ywc}. This gauge-equivariant ``convolution'' can be written explicitly as the recursion
\begin{equation}\label{eq:PTconv}
        V^{(i+1)}_\mu  = V^{(i)}_\mu + \sum_{ \substack{\rho \neq \mu,\\ \ell}} \eta^{\ell}_{i,\rho} ( R^{\ell}_{\mu\rho}(V^{(i)}) +  L^{\ell}_{\mu\rho}(V^{(i)}) ) \ ,
\end{equation}
where
\begin{equation}\label{eq:V0def}
     V_\mu^{(0)}(x) = \begin{cases}
        U_\mu(x) & \text{$U_\mu(x)$ is frozen,} \\
        0 & \text{$U_\mu(x)$ is active,}
    \end{cases}
\end{equation}
$\eta^{\ell}_{i, \rho}$ are trainable coefficients, and $L^\ell$ and $R^\ell$ label generic staple-like objects that transform in the same way as the gauge links.
Here we use two explicit choices, $R_{\mu \nu}^1 = (S_{x,\mu \nu}^R)^\dagger$ in \Cref{eq:staples} and $R_{\mu \nu}^2 = W_{x,\mu \nu}^R  U_\mu $, and similarly for $L_{\mu \nu}^{\ell}$; see \Cref{fig:archsketch}.
Note that in \Cref{eq:PTconv}, these objects are computed using the variables $V^{(i)}$.
After iterating, $V^{(i)}$ is not an element of the gauge group, but this is not important since ultimately there is a projection to the algebra to construct $G_\mu$ in \Cref{eq:Gmu}.

The iterative procedure in \Cref{eq:PTconv} can be used to construct expressive residual layers.
After applying $n_{\rm pt}$ iterations of \Cref{eq:PTconv} to \Cref{eq:V0def}, the resulting values of $V^{(n_{\rm pt})}$ can be used to construct the quantity $g_x(V^{(n_{\rm pt})}, U_\mu(x_a))$ that enters in the transformation of the residual layer defined in \Cref{eq:reslayer}.
Specifically, the convoluted frozen links, $V^{(n_{\rm pt})}$, are used to construct the staples in \Cref{eq:WRandWL} instead of $U_f$. 

\section{Example applications}\label{sec:demos}

Physics contexts in which derivatives of the form of \Cref{eq:derOalpha} arise are ubiquitous; here we discuss three examples.
First, derivatives with respect to the gauge coupling $\beta$ can be used to constrain continuum extrapolations.
Second, matrix elements may be computed using Feynman-Hellmann techniques, where derivatives with respect to action parameters correspond to single insertions of the corresponding operator.
Second-order derivatives using Feynman-Hellmann also access physically relevant processes, e.g.,~Compton scattering.
Third, derivatives with respect to the quark mass can be employed to constrain chiral extrapolations or in 
calculations of e.g.,~sigma terms.
This section presents numerical demonstrations using flows to improve estimates of these three kinds of derivatives.

The flow models used in these applications are summarized in \Cref{tab:modelcont}.
All flow models have been optimized using path gradients~\cite{vaitl2022gradients} as described in Ref.~\cite{Abbott:2023thq}.
Gauge field samples for both training and evaluation are obtained using standard Markov Chain Monte Carlo methods, specifically the (pseudo-)heatbath algorithm with overrelaxation~\cite{Creutz:1980zw, Cabibbo:1982zn, Kennedy:1985nu, Brown:1987rra, Adler:1987ce} for Yang-Mills theory and the Hybrid/Hamiltonian Monte Carlo~\cite{DUANE1987216} (HMC) algorithm for QCD.

\subsection{Continuum limit of gauge theories}\label{subsec:continuum}

One application in lattice QCD for flow-correlated ensembles is in taking the continuum limit.
For a numerical demonstration, we consider gradient flow scales.

We use the pure-gauge SU(3) theory, with action
\begin{equation}\label{eq:Spg}
   S_g(U)= -\frac{\beta}{N_c} \text{Tr Re } \sum_{\mu >\nu} U_{\mu\nu} \ ,
\end{equation}
where $\beta$ is the inverse squared bare gauge coupling and $U_{\mu\nu}$ is the plaquette.
The continuum limit of lattice spacing $a \to 0$ corresponds to $\beta \to \infty$.

One class of observables is obtained by using the gradient flow; in particular, a scale $t_c$ can be defined implicitly from
\begin{equation}\label{eq:tcdef}
    \langle t^2 E(t) \rangle \rvert_{t = t_c} = c \ ,
\end{equation}
where $c$ is a numerical constant, and $E(t)$ is the energy density 
at flow time $t$, for which we use the plaquette definition; see Eq.~(3.1) in Ref.~\cite{Luscher:2010iy}.
The choice $c=0.3$ defines the scale $t_{0.3}$, often referred to as ``$t_0$''.
One can compute the ratio of two gradient flow scales $t_{0.3}/t_{0.35}$, which can be related to the ratio of the the strong coupling at two different energy scales~\cite{Luscher:2010iy}.
The continuum limit of this quantity takes the form
\begin{equation}
    \frac{t_{0.3}}{t_{0.35}}\bigg \rvert_{\rm lat} =   \frac{t_{0.3}}{t_{0.35}}\bigg\rvert_{\rm cont}  + k_1 \frac{a^2}{t_{0.3}} + \cdots \ ,
\end{equation}
where $k_1$ is a dimensionless constant, the ellipsis indicates higher orders in $a^2$, the subscripts ``lat'' and ``cont'' refer to finite-$a$ and continuum values, and discretization effects are parameterized by powers of $a^2/t_{0.3}$.

The standard approach for performing a continuum extrapolation in lattice QCD relies on computing the desired quantity at several different lattice spacings using independent ensembles and extrapolating.
This method can be improved by additional constraints on such an extrapolation in the form of derivatives
\begin{equation}\label{eq:kder}
    k(a^2) = \frac{d\left( t_{0.3}/t_{0.35} \right)}{d (a^2/t_{0.3})} = k_1 + O(a^2) \ .
\end{equation}
Without generating more ensembles, this derivative can be computed using finite differences combined with $\epsilon$ reweighting or with flows to nearby values of the lattice spacing, or equivalently, values of the bare gauge coupling $\beta$:
\begin{equation}\label{eqn:cont-limit-target}
    k(a^2) \simeq \frac{ \frac{t_{0.3}}{t_{0.35}}\big \rvert_{\beta+\Delta \beta} - \frac{t_{0.3}}{t_{0.35}}\big \rvert_{\beta}  }{
    \frac{a^2}{t_{0.3}}\big \rvert_{\beta+\Delta \beta} - \frac{a^2}{t_{0.3}}\big \rvert_{\beta}   } \ .
\end{equation}
Note that the gradient flow scales $t_c$ are derived quantities, so we use the notation ``$\rvert_{\beta}$'' to indicate that they have been computed in a theory with the given $\beta$.

To demonstrate the advantage gained by using flows, we compute \Cref{eqn:cont-limit-target} using $\epsilon$ reweighting (\Cref{eq:eps-reweighting-def}) and the flowed approach (\Cref{eq:flow-reweighting-def}) and compare.
For this test, we use 96k configurations at $\beta=6.02$ on volume $L^4 = 16^4$.
For $\epsilon$ reweighting, we use a step of $\Delta \beta = 0.001$, leading to an ESS of 96\% on this ensemble.
For the flowed approach, we use Model A of \Cref{tab:modelcont}, which maps from $\beta=6.02$ to $\beta=6.03$, that is $\Delta \beta=0.01$.
This model achieves an ESS of $67\%$, which is significantly higher than direct reweighting, which has an ESS of 2\% at the same target parameters.
Using these approaches, we find
\begin{equation}\label{eq:numdemocont}
\begin{aligned}
      \text{Flow: }  k(a^2) &= -0.0167(41) \ ,\\
      \text{$\epsilon$ reweighting: }  k(a^2) &= -0.0208(63) \ ,
\end{aligned}
\end{equation}
that is, the statistical uncertainly using $\epsilon$ reweighting is 50\% larger than that obtained with flows.
In other words, one needs about $2.4 \times$ fewer samples using the flow method as compared with $\epsilon$ reweighting to achieve the same statistical uncertainty.  

\begin{figure}
    \includegraphics[width=8cm]{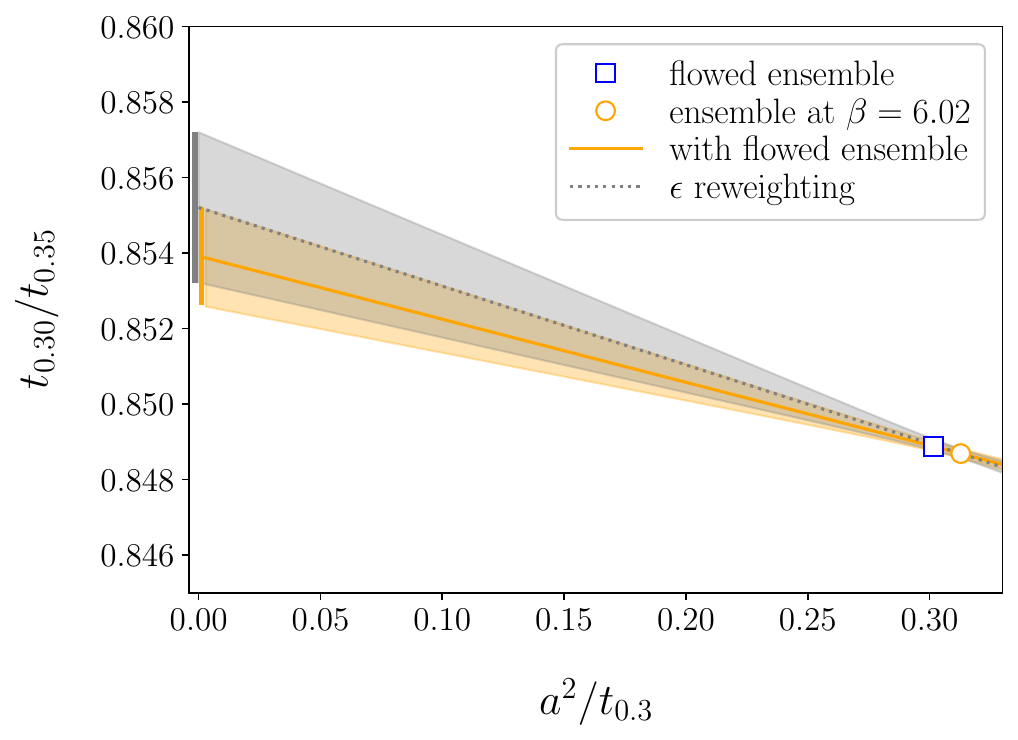}
    \caption{Continuum extrapolation of the ratio of two gradient flow scales $t_{0.3}/t_{0.35}$, using the quantity in the numerator to set the scale. Two methods are shown: $\epsilon$ reweighting (dotted grey line), and using a flowed ensemble (solid orange line). Statistical uncertainties are displayed as bands.}
    \label{fig:t0cont}
\end{figure}

Assuming that cutoff effects are already in the linear regime at this value of the lattice spacing, one can use this procedure to perform a simple continuum extrapolation of the ratio of flow scales. 
The continuum-extrapolated results show the same hierarchy of uncertainties as in \Cref{eq:numdemocont}:
\begin{equation}
\begin{aligned}
      \text{Flow: } t_{0.3}/t_{0.35}\rvert_{\rm cont} &= 0.8539(13) \ ,\\
      \text{$\epsilon$ reweighting: }  t_{0.3}/t_{0.35}\rvert_{\rm cont} &=0.8552(20) \ .
\end{aligned}
\end{equation}
These results are shown in \Cref{fig:t0cont} for the two methods.

\begin{table*}
    \begin{ruledtabular}
    \begin{tabular}{cccccccc}
    Model & Prior type & Parameters & Target type & Parameters & Train ESS & Eval. vol. & ESS \\ \hline
    A & Pure Gauge SU(3)& $\beta=6.02$ & Pure Gauge SU(3)& $\beta=6.03$ & 99.72\% & $16^4$ & $67 \%$  \\
    B1 & Pure Gauge SU(3)& $\beta=6.00$ & Feynman-Hellmann & $\beta=6.00, \lambda=+0.01$ & $99.4\%$ & $16 \times 8^3$ & $84 \%$ \\
    B2 & Pure Gauge SU(3)& $\beta=6.00$ & Feynman-Hellmann & $\beta=6.00, \lambda=-0.01$ & $99.4\%$  & $16 \times 8^3$ & $84 \%$ \\
    C & $N_f=2$ QCD & $\beta=5.60, \kappa=0.153$ &  $N_f=2$ QCD & $\beta=5.60, \kappa=0.1545$ & 99.2\% & $8^4$ & 48\%
    \end{tabular}
    \end{ruledtabular}
    \caption{Summary of flow models used in this work. All flow models have been trained on a hypercubic lattice volume of size $4^4$, while the evaluation lattice volume at which the flows are used (Eval. vol.) is given explicitly in the table.}
    \label{tab:modelcont}
\end{table*}

\subsection{Hadron structure with Feynman-Hellmann techniques}\label{subsec:FH}

Another promising application of machine-learned flows is in the calculation of matrix elements via the Feynman-Hellmann (FH) approach---see Refs.~\cite{QCDSF:2012mkm, Batelaan:2023jqp, QCDSFUKQCDCSSM:2022ncb, Hannaford-Gunn:2022lez} for recent applications.
In this framework, a matrix element
\begin{equation}
    T_h = \braket{ h | \mathcal O | h} \ ,
\end{equation}
where $h$ is a stable hadron at rest and $\mathcal{O}$ is the operator of interest projected to zero momentum, is computed by taking derivatives with respect to a parameter in the action.
Specifically, adding the operator to the action as
\begin{equation}
    S \to S_\lambda = S + \lambda \mathcal O \ ,
\end{equation}
the matrix element can be obtained as
\begin{equation}
    T_h = \frac{1}{2 M_h} \frac{d M_h}{d\lambda} \bigg \rvert_{\lambda \to 0} \ ,
\end{equation}
where $M_h$ is the hadron mass.
In practice, this can be estimated using a finite-difference approximation of the derivative, e.g.,
\begin{equation}\label{eq:finitediffFH}
    T_h = \frac{1}{2 M_h(0)} \frac{ M_h(+\lambda) -  M_h(-\lambda)   }{2 \lambda} + O(\lambda^2) \ ,
\end{equation}
Other improved finite-difference approximations or modeling-based approaches may also be used to better control the $O(\lambda^2)$ bias. 
As a numerical demonstration, we consider a Feynman-Hellmann calculation of the gluon momentum fraction of the pion in the quenched approximation of lattice QCD, similar to Ref.~\cite{QCDSF:2012mkm}.
In this case the operator $\mathcal O$ may be defined as
\begin{equation}
    \mathcal O = -\frac{\beta}{N_c} \text{Tr Re }\left( \sum_{i} U_{i0}  - \sum_{i<j}  U_{ij} \right) \ ,
\end{equation}
where $i,j \in (1,2,3)$, which is a discretization of the Energy-Momentum-Tensor (EMT). The matrix element can then be related to the gluon momentum fraction of the hadron $\langle x \rangle_g$ by
\begin{equation}\label{eq:mastereqFH}
  \frac{d M_h}{d\lambda}   \bigg \rvert_{\lambda \to 0} = -\frac{3M_h}{2} \langle x \rangle_g^{\rm latt} \ ,
\end{equation}
where the superscript ``$\text{latt}$'' emphasizes that it is a bare matrix element.
When adding this operator to the gauge action with a small parameter $\lambda$, the full action can be seen as an anisotropic action with different couplings for the temporal and spatial plaquettes:
\begin{equation}
\begin{aligned}
    S_\lambda = &-\frac{\beta}{N_c} (1+\lambda) \Re \Tr \sum_{i} U_{i0} \\
    &-\frac{\beta}{N_c} (1-\lambda) \Re \Tr \sum_{i<j} U_{ij} \ . 
\end{aligned}
\end{equation}
It is therefore possible to use flow transformations to map from the isotropic pure gauge action at $\lambda=0$ to non-zero values of $\lambda$.
This target is referred to as ``Feynman-Hellmann'' in \Cref{tab:modelcont}.

We test the flowed approach by computing the difference in \Cref{eq:finitediffFH} using an ensemble generated at $\lambda=0$ and flowed to non-zero $\pm \lambda$ values.  The choice $\lambda=0.01$ is small enough that $O(\lambda^2)$ discretization artifacts in the derivative are negligible; compare to the results in Ref.~\cite{QCDSF:2012mkm}.
We train two flows, B1 and B2 in \Cref{tab:modelcont}, which achieve an ESS of 84\% at the evaluation volume, cf.\ the direct reweighting ESS of around 2\% at the same values of $\lambda$.
The target parameters are matched to Ref.~\cite{QCDSF:2012mkm}, albeit at a smaller volume.
The value of $\beta=6$ corresponds to a lattice spacing of $a \simeq 0.09$ fm (using the Sommer radius to set the scale~\cite{Durr:2006ky}), and the hopping parameter $\kappa$ in the quenched Dirac operator---related to the bare quark mass as $\kappa = 1/(2m_0 +4)$---is taken to be $\kappa=0.132$.
The lattice spatial and temporal extent are $L=8$ and $T=16$, such that $M_\pi L > 4$.
For the purpose of this demonstration, we approximate the pion masses using the effective mass at the center of the lattice,
\begin{equation}
   \cosh aM_\pi = \frac{C_\pi (T/2+1) + C_\pi (T/2-1)}{2 C_\pi (T/2)} \ ,
\end{equation}
where $C_\pi(t)$ is the pion correlator. In physical units, $M_\pi \simeq 1.2$ GeV.

\begin{figure*}
     \centering
     \subfloat[\label{fig:FHmpi}]{%
     \includegraphics[width=0.49\textwidth]{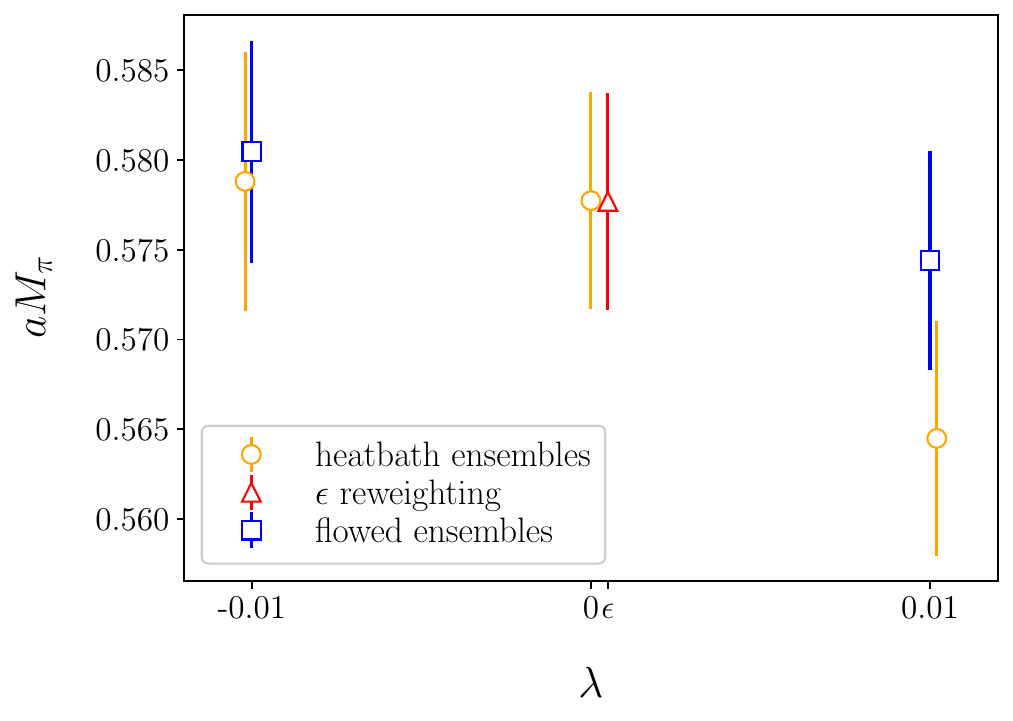}
    }
    \hfill
    \subfloat[\label{fig:xmom}]{%
     \includegraphics[width=0.49\textwidth]{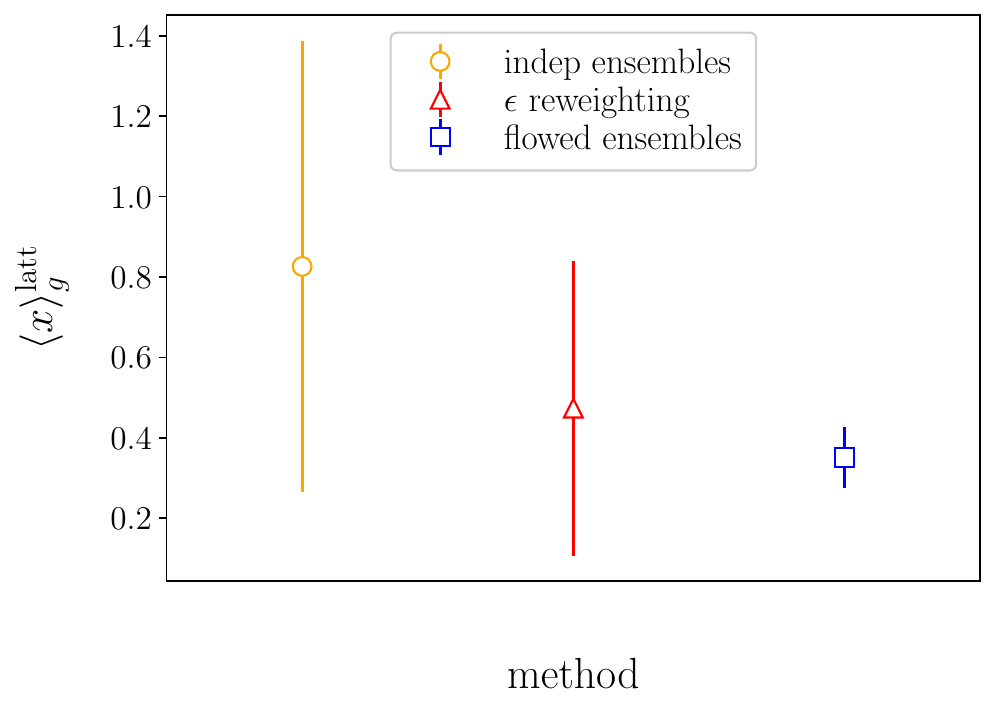}
    }
    \caption{(a) Pion mass in lattice units as a function of the coupling to the gluonic energy-momentum tensor $\lambda$.
    Marker shapes denote how the ensembles were obtained: orange circles for heatbath ensembles at fixed values of $\lambda$, blue squares for ensembles flowed from $\lambda=0$, and red triangles when using configurations generated at $\lambda=0$ and reweighted to ${\lambda=\epsilon=10^{-4}}$.
    The pion mass is evaluated in quenched lattice QCD at $\beta=6.0$, $\kappa=0.132$, $L=8$ and $T=16$.
    (b) Bare gluon momentum fraction of the pion from \Cref{eq:mastereqFH} using a finite-difference approximation computed using the three different methods: independent heatbath ensembles, $\epsilon$ reweighting, and correlated flowed ensembles.}
    \label{fig:ESS}
\end{figure*}

For evaluation, 14k gauge-field configurations are generated using 1 heatbath step with 5 overrelaxation steps between measurements for each independent ensemble.
Correlation functions are measured with four smeared sources per configuration with point sinks, using Chroma~\cite{Edwards:2004sx}.
The pion mass as a function of $\lambda$ is shown in \Cref{fig:FHmpi}, as determined using $\epsilon$ reweighting, independent ensembles, and flowed ensembles.
Since the flow model quality at the volume of interest is very high, uncertainties in the observables computed on flowed ensembles are very similar to those computed using ensembles generated with heatbath.

The physical quantity of interest, $\langle x \rangle^{\rm latt}_g$, depends on the difference between the pion mass determined at different values of $\lambda$.
When this difference is computed using independent ensembles, statistical uncertainties add in the usual way, and the error in the correlated difference is larger than that of each $M_\pi(\lambda)$ estimate.
In contrast, for flowed ensembles or $\epsilon$ reweighting, cancellations of correlated fluctuations significantly reduce the variances.
This can be seen in \Cref{fig:xmom}, which shows $ \langle x \rangle_g^{\rm latt}$ computed following the different methods outlined in \Cref{sec:corrflows}.
The use of flowed ensembles reduces the uncertainty by a factor of $\simeq 7$ with respect to independent ensembles, and $\simeq 5$ with respect to $\epsilon$ reweighting (using $\lambda=10^{-4}$ with an $\text{ESS}$ of 99.93\%).
Thus, incorporating flows into this calculation leads to a reduction of more than $20\times$
in the number of configurations necessary to achieve the same statistical error.

It is also possible to compute the second derivative of $M_\pi$ with respect to $\lambda$, which can be approximated as
\begin{equation}
    \frac{d^2M_\pi}{d \lambda^2} \bigg \rvert_{\lambda=0} \simeq \frac{ M_h(+\lambda) + M_h(-\lambda) -2  M_h(0)}{  \lambda^2} \ .
\end{equation}
While for the particular case of the gluon energy-momentum tensor this derivative is not physically relevant, second derivatives are related to matrix elements of two-current insertions---see for instance Compton scattering applications~\cite{Can:2020sxc, CSSMQCDSFUKQCD:2022fzy}.
Using the same three methods as for the first derivative, we find:
\begin{equation}
\begin{aligned}
      \text{Flow: } \frac{d^2M_\pi}{d \lambda^2} \bigg \rvert_{\lambda=0} &= -6(15) \ ,\\
      \text{$\epsilon$ reweighting: }  \frac{d^2M_\pi}{d \lambda^2} \bigg \rvert_{\lambda=0} &=-140(110) \ . \\
      \text{Indep. ens.: }  \frac{d^2M_\pi}{d \lambda^2} \bigg \rvert_{\lambda=0} &=-120(150) \ .
\end{aligned}
\end{equation}
All the determinations yield numbers that are zero within two standard deviations, but the relative magnitude of the uncertainties can nevertheless be used to assess the advantage of the flowed approach.
In particular, for the second derivative, the error reduction when using flows is larger than for the case of the first derivative, a factor of $7-10$ smaller than that obtained using $\epsilon$ reweighting or independent ensembles.
This, in turn, leads to requiring one to two orders of magnitude fewer configurations to achieve some target statistical precision. 

\subsection{Mass dependence of QCD observables}\label{subsec:mass}

As a third example, we compute derivatives with respect to the quark mass in QCD with $N_f=2$ unimproved Wilson fermions.
As a simple demonstration, we work directly with the action including the exact fermion determinant,
\begin{equation}\label{eq:Sexactdet}
    S(U) = S_g(U) - \log \det D_w[U] D^\dagger_w[U] \ ,
\end{equation}
where $S_g(U)$ is the plaquette gauge action and $D_w$ is the discrete standard Wilson operator.
The quark mass enters in the action via the hopping parameter $\kappa$. This target is referred to as ``$N_f=2$ QCD'' in \Cref{tab:modelcont}.

\begin{figure}
    \includegraphics[width=\linewidth]{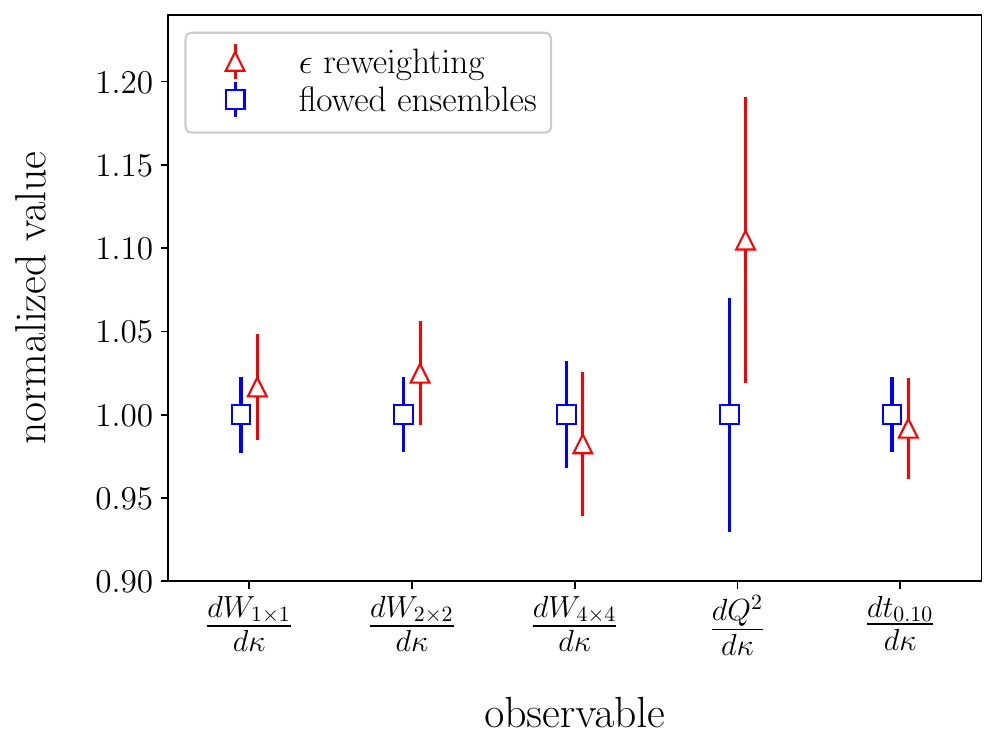}
    \caption{Illustration of the error reduction in derivatives of observables with respect to the action parameter $\kappa$.
    $W_{n \times n}$ is the average square Wilson loop of size $n$, $Q^2$ is the squared topological charge defined via the gradient flow, and $t_c$ labels gradient flow scales, as in \Cref{eq:tcdef}.
    The y-axis shows the values of the observables and their statistical errors normalized to the value obtained with flows.
    Results that incorporate flows are shown as blue squares, while the errors with $\epsilon$ reweighting are denoted by red triangles.}
    \label{fig:demoqcd}
\end{figure}

For this test, we compute the derivative of some simple observables (generically labelled as $X$) with respect to $\kappa$, approximated via finite differences:
\begin{equation}
    \frac{d X}{d \kappa} \simeq \frac{X(\kappa_2) - X(\kappa_1)}{\kappa_2-\kappa_1} \ .
\end{equation}
Depending on the observable, such derivatives can be useful, e.g., to extract sigma terms or to constrain chiral extrapolations. 
Here we specifically consider the average plaquette, the squared topological charge measured using the gradient flow at flow time $t/a^2=2$, and gradient flow scales $t_c$.

\begin{figure*}[t!]
    \includegraphics[width=10.28cm]{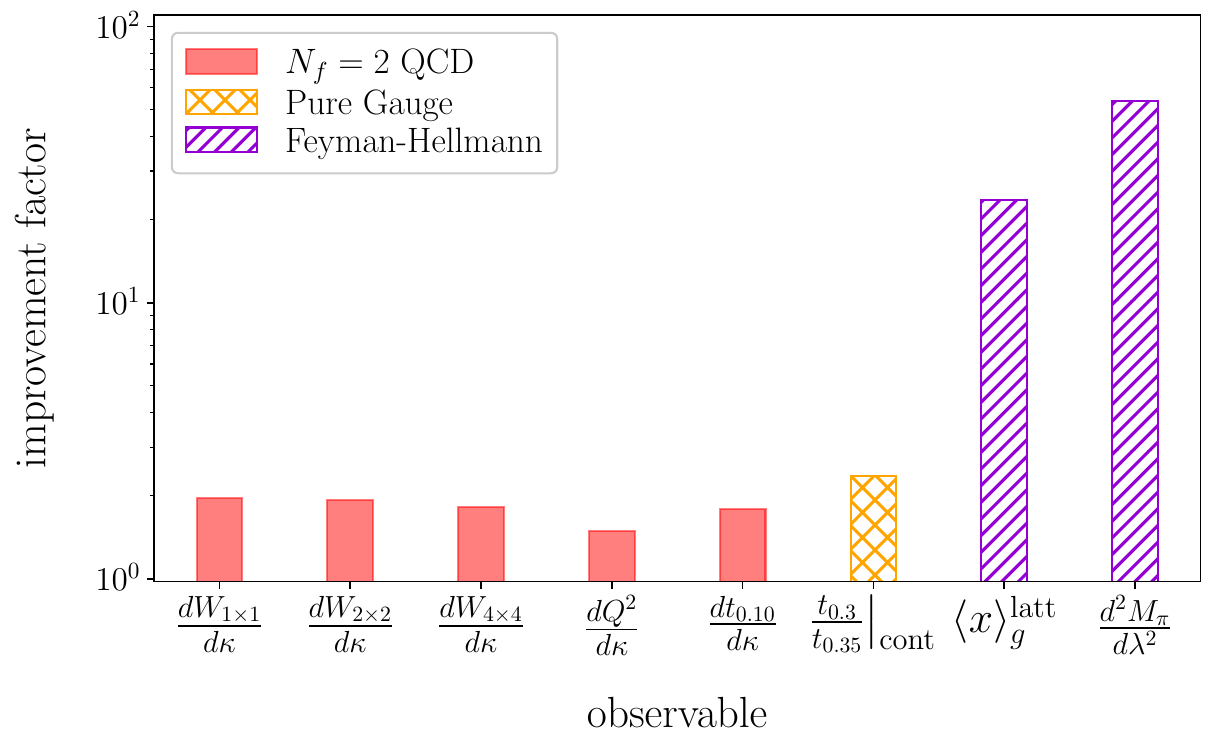}
    \caption{Summary of the variance reduction in observables computed from derivatives with respect to the action parameters when using flows compared with $\epsilon$ reweighting.
    The improvement factor is defined as the ratio of variances of the observables computed with $\epsilon$ reweighting over flows.
    The label ``$N_f=2$ QCD'' denotes derivatives of observables with respect to $\kappa$ in two-flavor QCD, the label ``Pure Gauge'' corresponds to the result for the continuum limit extrapolation of gradient flow scales in the pure gauge theory, and the label ``Feynman-Hellmann'' indicates observables computed using the Feynman-Hellmann approach in quenched QCD.}
\label{fig:improvement}
\end{figure*}

We train a flow to map configurations from $\kappa=0.1530$ to $\kappa=0.1545$ at $\beta=5.6$ (Model C in \Cref{tab:modelcont}).
Such parameters are close to those in Ref.~\cite{Gupta:1991sn}.
9k configurations are generated using standard HMC with pseudofermions.
Note, however, the reweighting factor and KL divergence for each configuration are computed with \Cref{eq:Sexactdet}; this is statistically consistent and introduces no approximations.
At the evaluation volume of $8^4$, the flow achieves ${\rm ESS} = 48\%$, which should be compared with the ${\rm ESS} = 28\%$ obtained using direct reweighting to the same target parameters. 

The results are given in \Cref{fig:demoqcd}, 
which compares the (normalized) values of several observables computed using the two methods, i.e., correlated flowed ensembles and $\epsilon$ reweighting (with $\Delta \kappa=3 \cdot 10^{-4}$ for an $\text{ESS}$ of 95\%). 
At these statistics and for these choices of $\kappa$, independent ensembles result in statistical errors $\gtrsim 2 \times$ larger than those attained with flows, and we do not display them.
In all cases, flows provide a variance reduction and the central values are consistent within a standard deviation with those obtained with independent ensembles, which indicates that systematic errors in the finite-difference approximation of the derivative are not significant in this example.
The error reduction varies between observables in the range $\sim 20\%-40\%$.
In particular, the largest reduction is seen for the $1\times1$ plaquette loop, while the smallest is seen for the topological charge.
Thus, depending on the observable of interest, one requires a factor of $1.5-2\times$ fewer configurations to obtain a comparable statistical error when using flows.

\section{Conclusion}\label{sec:conclu}

In this work, we present the application of machine-learned flows to the computation of observables involving derivatives.
Specifically, we use flows to map ensembles between distributions defined by different parameters in the lattice action.
By exploiting correlated cancellations of uncertainties between these ensembles, this application has the potential to provide a 
computational advantage in the evaluation of finite-difference approximations of derivatives. 

To illustrate this idea, we showcase three numerical demonstrations in the context of lattice QCD: continuum limit extrapolations, matrix elements using the Feynman-Hellmann approach, and the mass dependence of observables.
In all cases, flows provide a reduction of variance, which implies that fewer configurations are needed to achieve the same statistical error.
The improvement factor for all demonstrations of this work, defined as the variance reduction in observables computed using flows with respect to $\epsilon$ reweighting, is summarized in \Cref{fig:improvement}.
These values are in the range of $1.5 \times$ for observables in QCD to more than $20\times$ for quantities in the Feynman-Hellmann approach.
With higher-quality flow models, these factors can be improved.

This comparison does not account for the differing costs of the different steps in each method, namely generating the initial ensemble with heatbath, applying the flow (in the flowed case), and measuring correlation functions.
Of course, the potential advantages of this approach depend sensitively on not only the model used, but on the particular application, the cost of evaluating observables, how autocorrelations are treated, and the precision goal. 
For a ballpark comparison, consider the results for the computation of matrix elements in the Feynman-Hellmann approach. 
In this application, the cost of applying the flow is comparable to the cost of measuring correlation functions, while the cost of a heatbath update is less by an order of magnitude.
This amounts to a factor of $\lesssim 3$ increase in computational cost to achieve a variance reduction by a factor of more than 20.
This constitutes a real computational advantage of approximately one order of magnitude, neglecting the costs of training. 
Given expected further improvements through the continued development of flow architectures, these results are promising.

This work focuses on target actions that only depend on the gauge fields, e.g., pure gauge SU(3), quenched QCD, and exact-determinant QCD.
To generalize these results to state-of-the-art lattice QCD scales, where the fermion determinant cannot be explicitly evaluated, one must combine these flows with pseudofermion flows for QCD, as explored in Refs.~\cite{Albergo:2021bna,Abbott:2022zhs,Abbott:2022hkm}. 

As flow model technology for lattice QCD continues to advance, applications of correlated ensembles could be extended to compute other interesting quantities, such as sigma terms of hadrons or observables in QED+QCD.
If the success seen in the proof-of-principle applications of this work can be achieved in such contexts, it holds the potential to drive substantial advances in the field.

\section*{Acknowledgements}

We thank Michael Albergo, Kyle Cranmer, and Ross Young for useful discussions. RA, DCH, FRL, PES, and JMU are supported in part by the U.S.\ Department of Energy, Office of Science, Office of Nuclear Physics, under grant Contract Number DE-SC0011090. PES is additionally supported by the U.S.\ DOE Early Career Award DE-SC0021006, by a NEC research award, and by the Carl G and Shirley Sontheimer Research Fund. FRL acknowledges support by the Mauricio and Carlota Botton Fellowship. GK was supported by the Swiss National Science Foundation (SNSF) under grant 200020\_200424. This manuscript has been authored by the Fermi Research Alliance, LLC under Contract No.~DE-AC02-07CH11359 with the U.S. Department of Energy, Office of Science, Office of High Energy Physics. This work is supported by the U.S.\ National Science Foundation under Cooperative Agreement PHY-2019786 (The NSF AI Institute for Artificial Intelligence and Fundamental Interactions, \url{http://iaifi.org/}) and is associated with an ALCF Aurora Early Science Program project, and used resources of the Argonne Leadership Computing Facility which is a DOE Office of Science User Facility supported under Contract DEAC02-06CH11357. The authors acknowledge the MIT SuperCloud and Lincoln Laboratory Supercomputing Center~\cite{reuther2018interactive} for providing HPC resources that have contributed to the research results reported within this paper. Numerical experiments and data analysis used PyTorch~\cite{paszke2019pytorch}, JAX~\cite{jax2018github}, Haiku~\cite{haiku2020github}, Horovod~\cite{sergeev2018horovod}, NumPy~\cite{harris2020array}, and SciPy~\cite{2020SciPy-NMeth}. Figures were produced using matplotlib~\cite{Hunter:2007}.

\appendix

\section{Details of models}\label{app:arch}

In this appendix, we provide some additional details of the models of this work and the scheme used to train them.
It is important to stress that the hyperparameters and training schemes of these models have not been fine-tuned to be optimal, but they suffice for the present demonstration.
It is therefore likely that the model quality can be increased with further training or simple modifications of the hyperparameters. 

The layers considered in this work use a ratio of polynomials
\begin{equation}
    f(x) = \frac{1}{1+2x}  \frac{a_0 + a_1 x }{b_0 + b_1 x }
\end{equation}
to construct $g_x$ in \Cref{eq:gxdef}, where $a_i$ and $b_i$ are trainable parameters.

All models have $n_{\rm pt}=6$, where $n_{\rm pt}$ is the number of iterations of \Cref{eq:PTconv} in each layer.
This choice has been found to be empirically better than lower values of $n_{\rm pt}$.
In models A, B1, and B2 we alternate the masking pattern between mod 2 or mod 4, since empirically this results in slight improvements compared to just using the mod 2 masking at the same computational cost (a mod 4 stack is computationally equivalent to two mod 2 stacks).
The model architectures are shown in \Cref{tab:modeldeets}. 

\begin{table*}
    \begin{ruledtabular}
    \begin{tabular}{ccccccc}
    Model &  Number of layers & Masking patterns  &  Number of params. & Gradient steps & Learning rate & Training batch size  \\ \hline
    A & 96 & (M2 + M4) $\times$ 4 & 16k &12000 & $ 10^{-4}$  & 2048 \\
    B1 & 72  & (M2 + M4) $\times$ 3 &12k &2100 & $ 10^{-3}$ & 512\\
    B2 & 72 & (M2 + M4) $\times$ 3 & 12k&2100 & $ 10^{-3}$ & 512\\ 
    C & 88 & M2 $\times $ 11  & 15k& 1700  &$1.5 \cdot 10^{-3}$ & 960
    \end{tabular}
    \end{ruledtabular}
    \caption{Additional details of the flow models of this work. ``M2'' and ``M4'' refer to a masking pattern modulo 2 or modulo 4, respectively, as described in \Cref{subsec:arch}.}
    \label{tab:modeldeets}
\end{table*}

The models are optimized by minimizing the reverse KL divergence:
\begin{equation}
D_{\mathrm{KL}, \mathrm{rev}} = \frac{1}{B} \sum_{i=1}^{B} [\log q(U_i)+S(U_i)]+ \text{const.},
\end{equation}
where the sum runs over the $B$ configurations in a batch and the (unknown) normalization constant need not be evaluated for optimization.
Samples from the prior distribution are generated using heatbath/overrelaxation (pure gauge) or HMC (QCD).
The training scheme consists of a constant learning rate for a fixed number of gradient steps. 
We use a constant batch size to train each model. Between gradient steps, each configuration in the batch is evolved independently using the corresponding update algorithm.
These details are summarised in \Cref{tab:modeldeets}.

In all cases, we use path gradients, which are implemented by computing the gradients for optimization using the path derivative rather than the total derivative:
\begin{equation}
    \frac{d \log q(U)}{d\theta} \to  \frac{\partial \log q(U)}{\partial U} \frac{d U}{d\theta}
\end{equation}
This reduces the variance of the gradients without changing their expectation.
See Ref.~\cite{vaitl2022gradients} for more details.

These models have been trained for different wall times: 10 days using 6 nodes with 8 NVIDIA A100 GPUs each for model A, 2 days using 2 nodes for models B1 and B2, and 2 days using 4 nodes for model C. Note that no attempts have been made to optimize either the training procedure nor implementation of the approach to reduce training times.

A sufficient condition to guarantee invertibility of the residual layers (Lipschitz condition) is
\begin{equation}\label{eq:lipschitz-lie-bound}
    ||g_x(V_1) - g_x(V_2)|| < ||V_1 - V_2||, \
\end{equation}
where  $||\cdot||$ denotes the matrix norm. This is not explicitly enforced in the transformations used in this work, but we have not detected any violations in trained models. 
See Appendix B of Ref.~\cite{Luscher:2009eq} for a discussion on the Lipschitz condition.

\bibliographystyle{utphys}
\bibliography{main}

\end{document}